\documentclass[pra,aps,reprint,a4paper,superscriptaddress,notitlepage,preprintnumbers,longbibliography,showkeys]{revtex4-1}
\usepackage{graphicx}
\usepackage{amsmath}
\usepackage{amssymb}
\usepackage{color,xcolor}
\usepackage[pdfstartview=FitH,linkbordercolor = white,colorlinks = true,linkcolor=blue,citecolor=blue,urlcolor=blue,urlcolor=blue]{hyperref}
\def\fisunical{Dip. di Fisica, Universit\`{a} della Calabria, Via P. Bucci, Cubo 30C, I-87036 Rende (CS), Italy}
\def\infnlnfcs{INFN, sezione LNF, Gruppo collegato di Cosenza, Cubo 31C, I-87036 Rende (CS), Italy}
\def\UAM{Dept. de Qu\'{\i}mica, Univ. Aut\'{o}noma de Madrid, C/ Fco. Tom\'{a}s y Valiente 7, E-28049 Madrid, Spain}
\def\AgI{$(4{\times}4)$}
\def\AgII{$(2\sqrt{3}{\times}2\sqrt{3}){\rm{R30}}^{\circ}$}
\def\AgIII{$(\sqrt{13}{\times}\sqrt{13}){\rm{R13.9}}^{\circ}$}

\def\Sio{$(1{\times}1)$}
\def\SiI{$(3{\times}3)$}
\def\SiII{$(\sqrt{7}{\times}\sqrt{7})_{\rm{I}}$}
\def\SiIII{$(\sqrt{7}{\times}\sqrt{7})_{\rm{II}}$}
\newcommand{\bra}[1]{\langle #1|}
\newcommand{\ket}[1]{|#1\rangle}

\newcommand{\msc}[1]{\text{\textsc{#1}}}
\newcommand{\Ignore}[1]{}
\newcommand{\im}{{\rm Im}}

\mathchardef\myhyp="2D
\def\fBZ{$1^{\rm{st}}{\rm{BZ}}$}
\begin{document}
\title{Interband $\pi$-like plasmon in silicene grown on silver}
\author{\firstname{A.}~\surname{Sindona}}
\email{antonello.sindona@fis.unical.it}
\affiliation{\fisunical}
\affiliation{\infnlnfcs}
\author{\firstname{A.}~\surname{Cupolillo}}
\email{anna.cupolillo@fis.unical.it}
\affiliation{\fisunical}
\affiliation{\infnlnfcs}
\author{\firstname{F.}~\surname{Alessandro}}
\affiliation{\fisunical}
\author{\firstname{M.}~\surname{Pisarra}}
\affiliation{\UAM}
\author{\firstname{D. C.}~\surname{Coello Fiallos}}
\affiliation{\fisunical}
\author{\firstname{S. M. }~\surname{Osman}}
\affiliation{\fisunical}
\author{\firstname{L. S.}~\surname{Caputi}}
\affiliation{\fisunical}
\begin{abstract}
Silicene, the two-dimensional allotrope of silicon, is predicted to exist in a low-buckled honeycomb lattice, characterized by semimetallic electronic bands with graphenelike energy-momentum dispersions around the Fermi level~(represented by touching Dirac cones).
Single layers of silicene are mostly synthesized by depositing silicon on top of silver, where, however, the different phases observed to date are so strongly hybridized with the substrate that not only the Dirac cones, but also the whole valence and conduction states of ideal silicene appear to be lost.
Here, we provide evidence that at least part of this semimetallic behavior is preserved by the coexistence of more silicene phases, epitaxially grown on Ag(111).
In particular, we combine electron energy loss spectroscopy and time-dependent density functional theory to characterize the low-energy plasmon of a multiphase-silicene/Ag(111) sample, prepared at controlled silicon coverage and growth temperature.
We find that this mode survives the interaction with the substrate, being perfectly matched with the $\pi$-like plasmon of ideal silicene.
We therefore suggest that the weakened interaction of multiphase silicene  with the substrate may provide a unique platform with the potential to develop novel applications based on two-dimensional silicon systems.
\end{abstract}
\keywords{
\href{https://doi.org/10.1103/PhysRevB.00.001400}{DOI:10.1103/PhysRevB.00.00140}\\
$^{\ast}$A. S. and $^{\dagger}$A. C. contributed equally}
\maketitle
\section{introduction}
Following the isolation of graphene sheets by mechanical exfoliation of its
parent crystal graphite~\cite{1}, enormous effort has been directed towards
two-dimensional~(2D) crystals made of group-IV elements other than
carbon.

A particularly noteworthy example is silicene, the silicon equivalent
of graphene with a natural compatibility with current semiconductor
technology~\cite{7,7b,7c}. First-principles calculations of the structural properties
report an intrinsic stability of a honeycomb arrangement of Si atoms in
slightly buckled form, with mixed $sp^2$-$sp^3$ hybridization~\cite{2,3,4}.
This system,
referred to as freestanding or  {\Sio}  silicene, presents a graphenelike,
semimetallic electronic structure characterized by linearly dispersing
$\pi$ and $\pi^*$ bands around the Fermi energy $E_F$, thus effectively
allowing the charge
carriers to mimic massless relativistic particles~\cite{8,9}. Unlike graphene,
freestanding silicene has a large spin-orbit coupling, which would make it
suitable for valley-spintronic applications~\cite{10,10b}, and an electrically,
magnetically, or chemically tunable band gap~\cite{11a,11b,11,11c,11d,11e}, which would be
crucial for engineering on-off current ratios and charge-carrier mobility in
silicene field-effect transistors.

On the practical side, silicenelike
nanostructures are synthesized by the epitaxial growth~\cite{14,13b} of silicon on
silver~\cite{12,6b,13,14cc,5,6,6c}  and a few other metal substrates~\cite{14b,14c}, very recently
including gold~\cite{47}. In particular, a number of well-ordered domains have
been observed on Ag(111), the formation and coexistence of which depend
on the substrate temperature, during silicon growth, and the silicon
deposition rate~\cite{15,16,17,17bb,17ccc,18,23}.

\begin{figure}[t]
\centering{\includegraphics[width=0.48\textwidth]{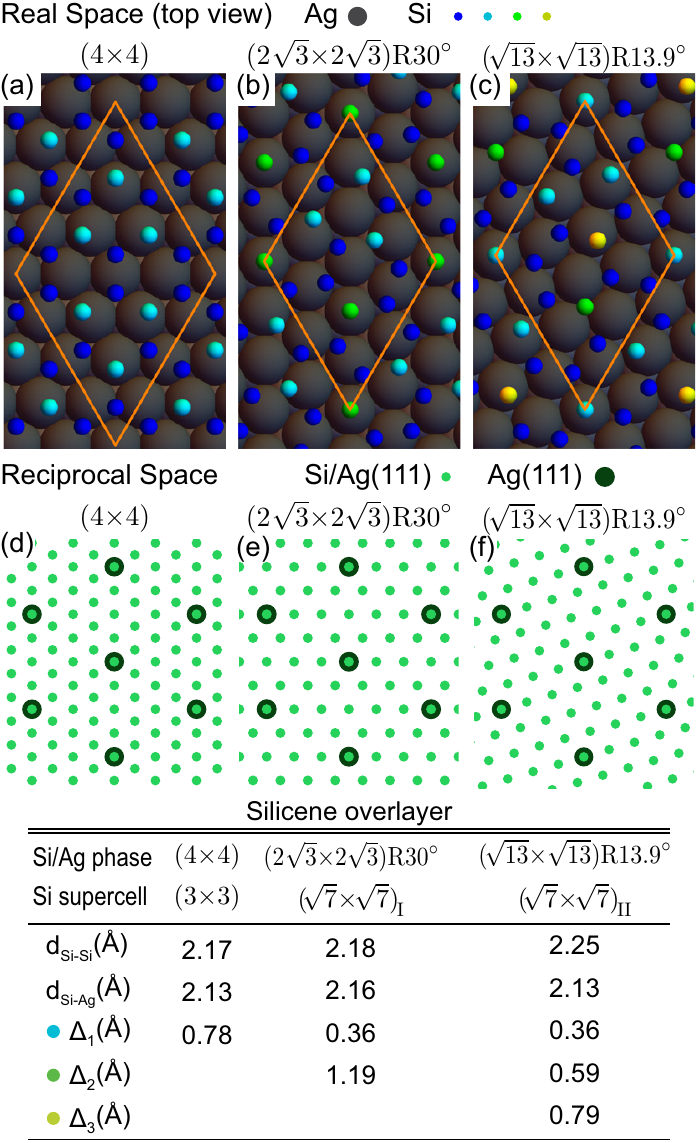}}
\vskip -6pt
\caption{Real and reciprocal spaces of the {\AgI}, {\AgII}, and one of
the two inequivalent {\AgIII} phases of silicene on silver~\cite{18,23,17ccc}.
The unit cells of the superstructures are delimited by orange lines. The main
geometric parameters of the silicene overlayers (commensurate supercell,
average in-plane bond length $d_{\rm{Si-Si}}$, minimum Si-Ag distance
$d_{\rm{Si-Ag}}$, and
out-of-plane atomic buckling $\Delta_i$) are derived from Refs.~\cite{27,28}.}
\label{F1phases}
\end{figure}

Most of these domains have a crystalline
morphology that is closely commensurate with the {\AgI}, {\AgII},
and {\AgIII} phases~\cite{17bb,18,23,20,21,24,26,26BB,26CC,26dd,26eeex,26eeey,22}
of Fig.~\ref{F1phases}. The different orientation,
out-of-plane atomic buckling, and lattice constant of the Si atoms in these
superstructures, here denoted {\SiI}, {\SiII}, and {\SiIII}, are refected in
markedly distinct electronic properties of the silicene overlayers that, even
without the supporting Ag substrate below, may or may not preserve the
Dirac cones of  {\Sio}  silicene~\cite{17ccc,17bb,27,28}. Other phases and more
complex scenarios, including silicon-induced faceting, are reported in the
literature~\cite{6c,20,17ccc,27,28yyy,28xxx}.

As for the interaction with Ag(111), silicene
seems to lose the unique properties of its {\Sio} form, as confrmed by
angle-resolved photoemission spectroscopy (ARPES) measurements
combined with density functional calculations~\cite{25,19,25bb,29,30,31}. Indeed, the initial
evidence of Dirac cones in the {\AgI} phase~\cite{6,23} has been reinterpreted
as originating from the s and p states of bulk Ag, or the strongly hybridized
sp states of Ag and Si~\cite{31}. More recent developments~\cite{32} suggest that a
hybridized Dirac-cone structure is unusually formed, as a result of the
aforementioned hybridization. Even in this latter case, however, the $\pi$-like
bands of freestanding silicene are destroyed by its interaction with Ag.
Thus, contrary to well-consolidated technologies of graphene growth on
copper, the transfer of silicene to other substrates appears impractical.

The intercalation of alkali-metal atoms between silicene and Ag(111) is
predicted to effectively restore the Dirac cones~\cite{32b}. Another possibility has
been indicated by changing the supporting substrate to Au(111)~\cite{47}.
Nonetheless, the major goal of silicene technology is still to grow a
honeycomblike structure of Si atoms with semimetallic, $\pi$-like electronic
features.

A different possible direction is outlined here, where, rather than
looking at the band structure of silicene on Ag(111), we consider exploring
the intrinsic plasmonics of the interface, i.e., the coherent charge-density
oscillations of its valence electrons (plasmons).
Specifically, we present
low-energy electron diffraction (LEED) and electron energy loss
spectroscopy (EELS) characterizations of two surface reconstructions of
silicene on Ag(111), being, respectively, given by the  {\AgII} and a
mixture of the {\AgI}, {\AgII}, and {\AgIII} superstructures~(Sec.~\ref{secI}).
We report a clear signature of the existence of $\pi$ and  $\pi^*$ states
of silicene in
the mixed phase, which reflects in a $\pi$-like plasmon structure detected
in the
energy loss (EL) spectrum of the system, alongside the
Ag plasmon~(Sec.~\ref{secI}).

The
measurements are consistent with time-dependent (TD) density functional
theory (DFT) (TDDFT) calculations of the EL function of {\Sio} silicene~\cite{33},
 showing a well-resolved loss peak at 1.75~eV for a momentum transfer
of the order of 10$^{-2}$~{\AA}$^{-1}$~(Sec.~\ref{secII}).
Furthermore, our results indicate that at least part
of the $\pi$ or  $\pi^*$ character of the first valence band (VB) or
first conduction
band (CB) of silicene grown on Ag(111) may be maintained under specific
geometric conditions~(Sec.~\ref{secII}).
On the other hand, the $\pi$-like mode is absent in each single pure phase of
the above-mentioned types, thus confirming that the semimetallic character
of freestanding silicene is lost in favor of a hybridized
band structure~(Sec.~\ref{secIII}).

\section{Preparation and characterization of the samples\label{secI}}
To begin, the experiments were carried out at room temperature in an
ultrahigh vacuum chamber (with a base pressure of 2${\times}$10$^{-10}$~Torr), using a
high-resolution electron energy loss (EEL) spectrometer, equipped with two
50-mm hemispherical defectors, as both monochromator and analyzer.

The
EL spectra were acquired with an incident electron beam, about 0.2 mm
wide, positioned at a fixed angle of 45$^{\circ}$ from the surface normal.
The kinetic
energy of the incident electrons was 40 eV.
Scattered electrons were
collected at an angle of 46$^{\circ}$ from the surface normal,
with an overall energy
 resolution of 20 meV and an angular acceptance of 2$^{\circ}$.

The silver substrate (a disk of about 8 mm diameter) was cleaned through
several cycles of sputtering with 1 keV Ar$^+$ ions, at an
operating pressure of 5${\times}$10$^-5$ Torr, followed by annealing at 500$^{\circ}$C until a
sharp LEED pattern showed the achievement of a clean and well-ordered
Ag(111) surface.

Silicon was deposited on Ag(111) by thermal evaporation
from a pure source, which was slab cut from a silicon wafer, clamped
between two electrodes, and heated up, by passing a direct current
thorough it, to obtain a stable vapor flux.
During deposition, the silicon source was placed in front of the silver
substrate, which was maintained at a constant temperature, controlled by a
thermocouple placed close to the sample. The pressure during growth was
less than 2${\times}$10$^-9$ Torr. The uptake of Si was monitored by x-ray
photoelectron spectroscopy and LEED.

\begin{figure}[t]
\centering{\includegraphics[width=0.499\textwidth]{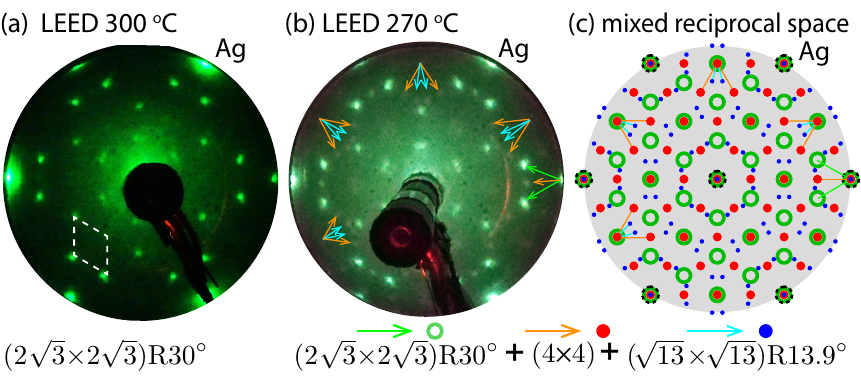}}
\vskip -6pt
\caption{LEED patterns of (a) the  {\AgII} phase and
(b) the {\AgI} + {\AgII} + {\AgIII} phase of silicene on Ag(111).
The coexistence
of multiple domains in (b) is attested by spots of different intensity, being
consistent with the mixed reciprocal space representation in (c). The latter
includes the reciprocal space points of Figs.~\ref{F1phases}(d)-\ref{F1phases}(f), plus the reciprocal
structure of the other {\AgIII} phase~\cite{18,23,17ccc} not reported in
Fig.~\ref{F1phases}.}
\label{leedx}
\end{figure}
Figure~\ref{leedx}(a) shows the diffraction
pattern acquired with the Ag substrate kept at 300$^{\circ}$C, during Si deposition.
The LEED spots from this first sample are consistent with the reciprocal
space points of Fig.~\ref{F1phases}(e), also reported in Fig.~\ref{leedx}(c). Accordingly, they are
readily interpreted as originating from a sufficiently pure  {\AgII} phase,
in agreement with several previous studies~\cite{15,16,26BB,26CC,26eeex}. On
the other hand, the LEED image of Fig.~\ref{leedx}(b) was acquired with the Ag
substrate kept at 270$^{\circ}$C, during Si deposition.

\begin{figure}[t]
\centering{\includegraphics[width=0.4\textwidth]{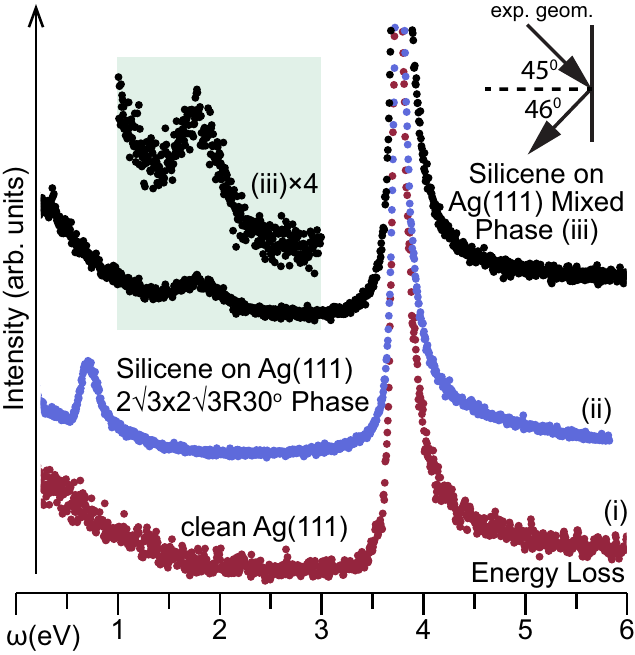}}
\vskip -6pt
\caption{EL spectra acquired, with a primary electron energy of 40 eV
and the experimental geometry shown in the inset, from:
(i) clean Ag(111)
(ii) silicene on Ag(111) in the pure {\AgII} phase and
(iii) silicene on Ag(111) in the mixed {\AgI}+{\AgII}+{\AgIII}
phase.}
\label{F2spec}
\end{figure}

This second sample is characterized by a mixture of the {\AgI},{\AgII},
and{\AgIII} phases, confirming that the growth conditions are very sensitive to the substrate temperature~\cite{16,17,17ccc}.
 The possible formation of recently isolated multilayer silicene can be ruled out, because it
would provide a different LEED pattern~\cite{25}.

Indeed, the LEED spots from the second sample are in line with the reciprocal
space points of the three phases, reported in Fig.~\ref{leedx}(c), where the silicene
superstructures are assumed to be perfectly matched with the Ag substrate.
Interestingly, a visual inspection of Figs.~\ref{leedx}(b) and~\ref{leedx}(c) suggests
a lattice mismatch of the {\SiII}  and {\SiIII}  overlayers of
$\sim$1$\%$ and $\sim$5$\%$, respectively.
Similar patterns, mostly concerning the {\AgI}and{\AgIII}phases,
have been discussed elsewhere~\cite{14,15,26eeex,26eeey,26dd}, nonetheless, our composite
structure, presenting a dominant  {\AgII} phase, which coexists with
the {\AgI}and{\AgIII}phases has yet to be reported.

The EELS measurements were performed with the samples directly transferred
in the ELL spectrometer, without breaking the vacuum, and kept at room
temperature with a pressure below 2${\times}$10$^{-10}$ Torr.
Figure~\ref{F2spec} shows the EL spectra, in the range of 0-6 eV, obtained from
(i) clean Ag(111), (ii) the pure  {\AgII} phase with the LEED pattern of
Fig.~\ref{leedx}(a), and (iii) the mixed phase with the LEED pattern of Fig.~\ref{leedx}(b).

A prominent feature at $\sim$3.8 eV, which corresponds to the surface plasmon of
Ag(111), dominates all spectra.
Besides, the Si/Ag(111) spectrum shows an additional loss at $\sim$0.7eV in the
pure phase and 1.75 eV in the mixed phase.
The excitation energy of the former mode suggests the presence of a
hybridized Si-Ag plasmon, whereas the latter mode resembles
the $\pi$-like plasmon of {\Sio} silicene~\cite{33}.

\section{Intrinsic plasmonics of the silicene systems\label{secII}}
\begin{figure}[t]
\centering{\includegraphics[width=0.495\textwidth]{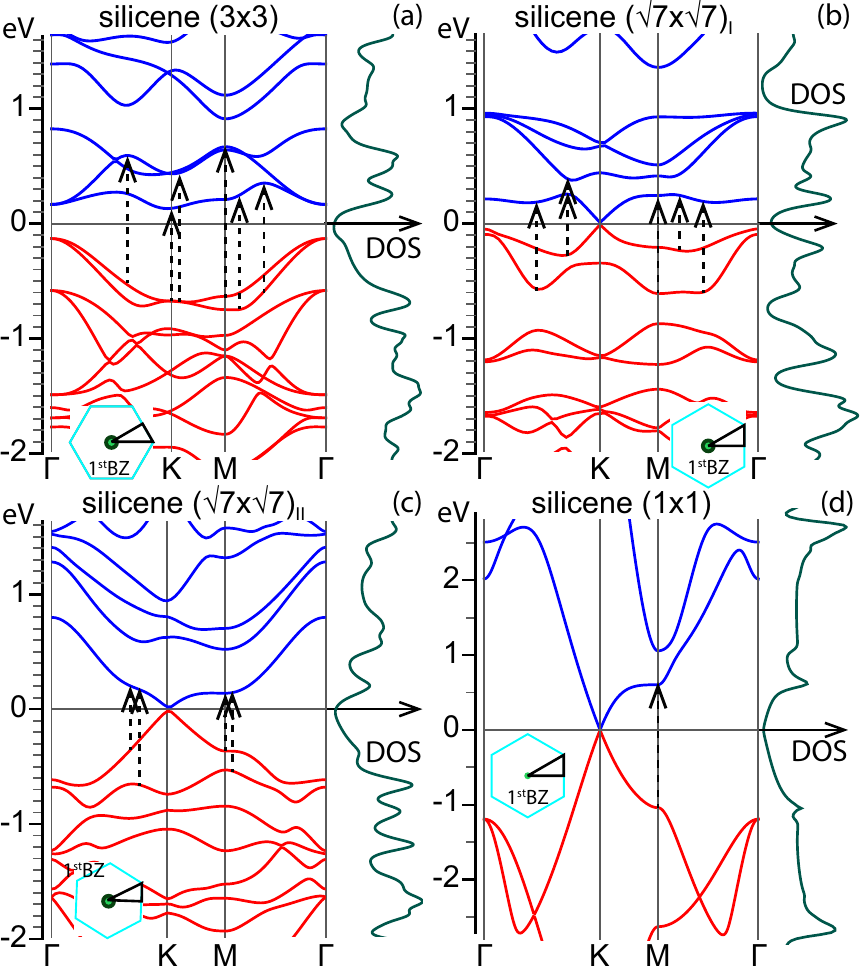}} \vskip -6pt
\caption{Band structure (with $E_F=0$) and DOS for {\SiI}, {\SiII}, {\SiIII},
and {\Sio} silicene.
The leading transitions concurring to the lower-energy loss peaks
of Fig.~\ref{F4comp} are represented by dashed arrow lines.
The ${\Gamma}{\rm{K}}{\rm{M}}{\Gamma}$ paths of the superstructures
are defined in the reciprocal spaces of Fig.~\ref{F1phases}(d)-(f).
}
\label{F3bands}
\end{figure}

To support the unprecedented observation outlined in the previous section,
we used TDDFT in the random phase approximation
(RPA) and computed the dielectric properties of {\Sio} silicene~\cite{33} in
comparison with the {\SiI}, {\SiII}, and {\SiIII} overlayers of silicene,
peeled from the silver substrate,
in the {\AgI}, {\AgII}, and {\AgIII} phases of Fig.~\ref{F1phases}.
These superstructures represent instructive examples of monolayer silicene,
which, without the Ag substrate below, would not~[Fig.~\ref{F3bands}(a)]
and would~[Fig.~\ref{F3bands}(b) and~\ref{F3bands}(c)] preserve the
Dirac-cone features~\cite{17ccc,17bb,27,28}.

As it is customary in TDDFT~\cite{34}, we started with a calculation of the
ground-state electronic properties of the silicene systems by DFT~\cite{35} in the
Kohn-Sham (KS) approach~\cite{37}, using the generalized gradient
approximation (GGA)~\cite{36a} and eliminating the core electrons by
suitable norm-conserving
pseudo-potentials~\cite{38}.
We selected a plane-wave~(PW) basis set, within an energy cutoff of 25~{hartree},
 to represent the KS electron states $\ket{\nu{\bf{k}}}$,
associated with the band energies $\varepsilon_{\nu{\bf{k}}}$, being labeled
by a band index $\nu $ and a wave vector ${\bf k}$ in the first
Brillouin zone~({\fBZ}).
The bulk-geometry, inherent to PW-DFT, was generated by replicating the
silicene slabs with an out-of-plane vacuum distance of $25$~{\AA}.

Geometry optimization was performed on unsupported {\Sio} silicene,
obtaining a lattice constant of $3.82{\,}{\rm{\AA}}$ and a characteristic
buckling of the AB type, specified by a buckling distance of
$0.45{\,}{\rm{\AA}}{\,}$~\cite{33}.
\Ignore{
 Further details on the electronic features of
the {\AgII} phase are provided in Sec. A of the
Supplemental Material~\cite{supX}.}

As for supported silicene, the substrate was simulated by four silver planes
with the morphologies of Fig.~\ref{F1phases},
starting from a set of optimized parameters found in Refs.~\cite{27,28}.
Then,  the positions of all atoms in the silicene and topmost Ag layers were
relaxed using a ${\Gamma}{\myhyp}{\rm{centered}}$
Monkhorst-Pack (MP) grid~\cite{39} of ${12}{\times}{12}{\times}{1}$
${\bf{k}}{\myhyp}{\rm{points}}$ in the {\fBZ},
with the occupied and the first few empty band-states of the systems.
Further details on the electronic features of the {\AgII} phase are provided
in Sec.~\ref{App1} of the Supplemental Material~\cite{supX}.

Single-point self-consistent runs were carried out  in the local density
approximation (LDA)~\cite{36} on a finer MP mesh
of ${60}{\times}{60}{\times}{1}$ ${\bf{k}}{\myhyp}{\rm{points}}$,
by removing the silver layers and keeping the same optimized positions for
the Si atoms as the geometry optimization step.
Subsequently, the converged electron densities were used, in non-self-consistent
runs, to obtain the eigensystem
$(\ket{\nu {\bf k}},\varepsilon _{\nu \mathbf{k}})$ on highly resolved
grids of  ${240}{\times}{240}{\times}{1}$ points, for {\Sio} silicene,
${90}{\times}{90}{\times}{1}$ points, for {\SiI} silicene,
and ${72}{\times}{72}{\times}{1}$ points for the other two silicene
superstructures~[{\SiII} and {\SiIII}].
A number of empty bands covering the KS energy spectrum up to 15 eV,
above $E_F$, was selected in all non-self-consistent cases, to accurately
describe the systems' plasmonics at energy losses below 6~eV.

The electronic structure and density of states~(DOS) of our DFT+LDA computations
are reported in Fig.~\ref{F3bands}, with the band energies represented along
the ${\Gamma}{\rm{K}}{\rm{M}}{\Gamma}$ paths of the corresponding
reciprocal spaces~(Fig.~\ref{F1phases}).
Here, we spot some peculiar features partly recognized in previous
studies~\cite{27,28,33}.
{\SiI} silicene has no Dirac cones at K and presents a direct gap at
${\Gamma}$ of ${\sim}0.3$~eV, between two couples of nearly degenerate
states of leading $\sigma$ and $\pi$ symmetries,
respectively~[Fig.~\ref{F3bands}(a)].
{\SiII}, and {\SiIII} silicene exhibit a quasilinear {$\pi$-like}  dispersion
in the first VB and first CB at $K$, i.e., a Dirac-cone structure with a tiny band
gap below ${\sim}0.02{\,}{\rm{eV}}$~[Fig.~\ref{F3bands}(b), (c)].
The second VB and CB of these superstructures are of dominant
$\sigma$ symmetry.
{\Sio} silicene has two quasimetallic bands of dominant
$\pi$ character~[Fig.~\ref{F3bands}(d)], whose dispersions
resemble the $\pi $ bands of graphene~\cite{33xx} on a reduced energy scale
and with a smaller Fermi velocity~\cite{33}.
Even in the latter case, the second VB and CB are dominantly $\sigma$, however,
a $\pi\leftrightarrow\sigma$ inversion has been detected at the crossing points
between the first and second VBs and the (avoided) crossing points between
the first and second CBs~\cite{33}.

\begin{figure}[!h]
\centering{\includegraphics[width=0.415\textwidth]{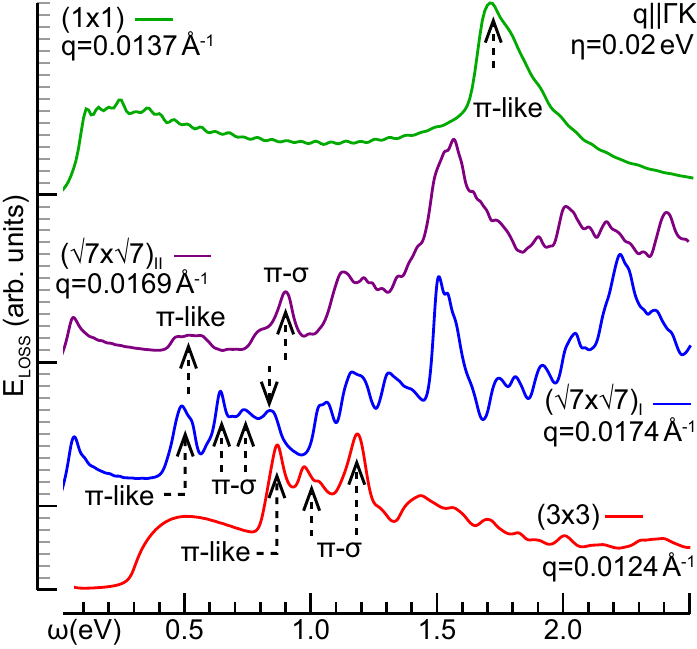}} \vskip -6pt
\caption{Theoretical loss function of {\SiI},  {\SiII},  {\SiIII}, and {\Sio}
silicene for $q{<}0.02{\,}{\rm{\AA }}^{-1}$
and ${\omega}{<}2.5{\,}{\rm{eV}}$, with the lower-energy peaks,
associated with transitions close to $E_F$, being represented
by dashed arrow lines (as ${\rm{in}}{\,}{\rm{Fig.}}{\,}$\ref{F3bands}).}
\label{F4comp}
\end{figure}

All these silicene systems have specific high intensity peaks in
the DOS that support electron excitations between the first or second VB and CB
near $E_F$, in a region of the {\fBZ} where the bands are quasiflat, and
may assist {$\pi$-like} and $\pi$-$\sigma$  plasmon oscillations.
The corresponding low-energy vertical transitions, which lead to small-momentum
EL-peaks, are indicated by dashed arrow lines in Figs.~\ref{F3bands}
and~\ref{F4comp}.

As a second step of the TDDFT framework, we computed the unperturbed
susceptibility of the KS electrons to an energy loss $\omega $ and momentum
 transfer ${\bf q}$ from the incident electron, provided by~\cite{40}
\begin{equation}
\chi _{\mathbf{G}\mathbf{G}^{\prime }}^{0}=\frac{2}{\Omega }\sum_{\mathbf{k}%
,\nu ,{\nu }^{\prime }}\frac{(f_{\nu \mathbf{k}}-f_{{\nu }^{\prime }\mathbf{k%
}+\mathbf{q}})\rho _{\nu {\nu }^{\prime }}^{\mathbf{k}\mathbf{q}}(\mathbf{G}%
)\,\rho _{\nu {\nu }^{\prime }}^{\mathbf{k}\mathbf{q}}(\mathbf{G}^{\prime
})^{\ast }}{\omega +\varepsilon _{\nu \mathbf{k}}-\varepsilon _{{\nu }%
^{\prime }\mathbf{k}+\mathbf{q}}+i\eta }. \label{AdlWi}
\end{equation}%
The latter (reported in Hartree atomic units) includes a factor of 2 for the
electron spin, the Fermi-Dirac occupations $f_{\nu \mathbf{k}}$
and $f_{\nu \mathbf{k}+\mathbf{q}}$~(evaluated at room temperature),
a lifetime broadening parameter $\eta $~(set to $0.02{\,}{\rm{eV}}$),
and the density-density correlation matrix elements
$\rho_{\nu {\nu }^{\prime }}^{\mathbf{k}\mathbf{q}}(\mathbf{G})
{=}\bra{\nu {\bf k}} e^{-i(\mathbf{q}+\mathbf{G})\cdot \mathbf{r}}
\ket{\nu'{\bf k}+{\bf q}}$.

The  full susceptibility was obtained using the central equation of TDDFT~\cite{42}:
\begin{equation}
\chi _{\mathbf{G}\mathbf{G}^{\prime }}{=}\chi _{\mathbf{G}\mathbf{G}^{\prime }}^{0}+(\chi ^{0}v\chi )_{\mathbf{G}\mathbf{G}^{\prime }}.
\end{equation}
Working the RPA, we represented the interaction matrix elements
in $\chi _{\mathbf{G}\mathbf{G}^{\prime }}$ by a truncated Coulomb potential,
which has been accurately designed to exclude the unphysical interaction between
the silicene slab replicas in a broad range of transferred momenta across
the whole {\fBZ}~\cite{43,44,45}.

Accordingly, the inverse permittivity was obtained
as
$
(\epsilon ^{-1})_{\mathbf{G}\mathbf{G}^{\prime }}{=}\delta _{\mathbf{G}%
\mathbf{G}^{\prime }}+(v\chi )_{\mathbf{G}\mathbf{G}^{\prime }}
$
and the EL
function was calculated from the macroscopic average $E_{\msc{loss}}{=}-\im[(\epsilon^{-1})_{{\bf 0}{\bf 0}}]$.
Nonlocal field effects~\cite{46} were included in $E_{\msc{loss}}$ through a
dimensional cut-off on the central equation of TDDFT, which we verified to converge
by reduction to a $61\times 61$ matrix equation, including the smallest
$\mathbf{G}$ vectors sorted in length from
$0{\,}{\rm{to}}{\,}{\sim}{\,}9.5{\,}{\rm{\AA }}^{-1}$.

\begin{figure}[!h]
\centering{\includegraphics[width=0.49\textwidth]{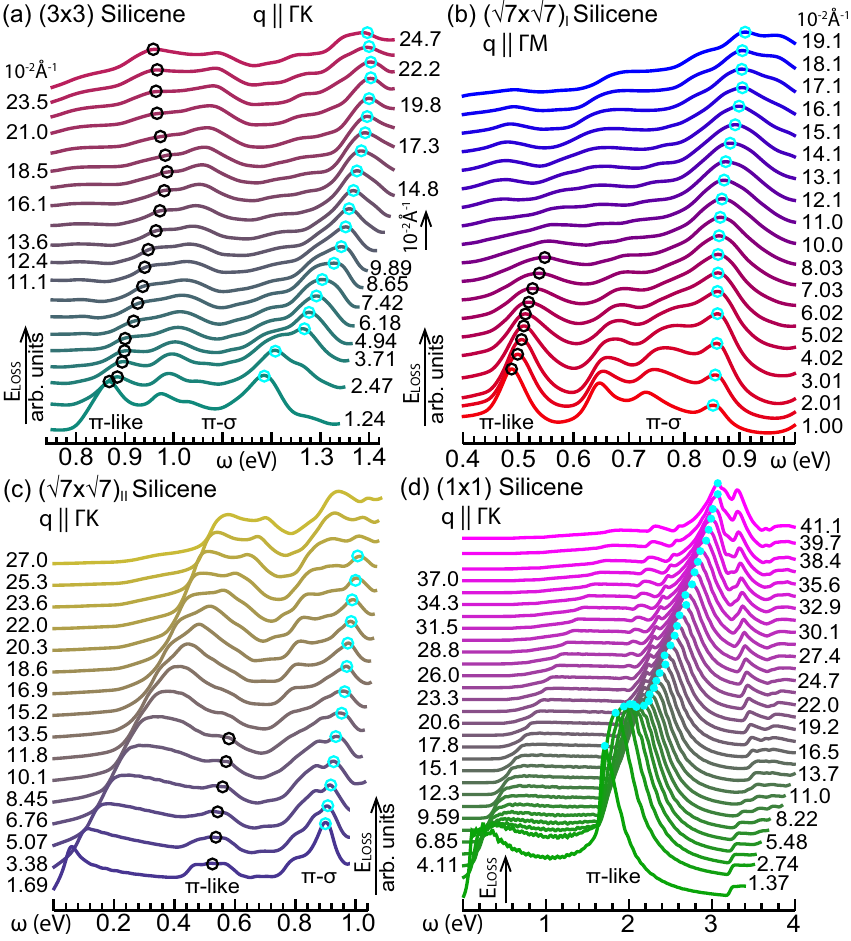}}
\vskip -6pt
\caption{Theoretical EL functions of (a) {\SiI}, (b) {\SiII}, (c) {\SiIII}, and (d) {\Sio}
silicene along the $\Gamma{K}$ [(a),(b),(d)] and $\Gamma{M}$ [(c)] directions of
the \fBZ{s} (insets of Fig.~\ref{F3bands}).}
\label{Fxdisp}
\end{figure}

In Fig.~\ref{F4comp}, we compare the theoretical loss spectra of our silicene systems,
at fixed small momentum transfers parallel to ${\Gamma}{\rm{K}}$ in an energy loss
range below $2.5{\,}{\rm{eV}}$.
The superstructures present a sequence of low-energy peaks associated with
quasivertical electron excitations between the first or second VB and CB,
of dominant $\pi$ or $\sigma$ character, given in Fig.~\ref{F3bands}(a)-(c).
Accordingly, the momentum-dependent investigation of Fig.~\ref{Fxdisp}(a)-(c)
let us identify two dispersive modes, which we respectively attribute to a {$\pi$-like}
plasmon, assisted by first-VB-to-first-CB transitions, and a $\pi$-$\sigma$ plasmon,
assisted by second-VB-to-first-CB and first-VB-to-second-CB transitions.
Much insight into the latter oscillation in {\SiII} silicene is provided
in Sec.~\ref{App2} of the Supplemental Material~\cite{supX}.

More importantly, we notice that there is no way that the EL functions
of the silicene superstructures can be combined to reproduce the experimental
loss spectrum of the mixed phase~[curve (iii) in Fig.~\ref{F2spec}].
On the contrary, {\Sio} silicene exhibits a broad primary {$\pi$-like}  plasmon
peak~[Figs.~\ref{F4comp} and~\ref{Fxdisp}(d)], consistent with the EELS
spectrum of the mixed phase, which is assisted by first-VB-to-first-CB transitions
around the M-point and superimposed to secondary one-electron excitation structures.
The detailed features of this mode have been carefully addressed in in Ref.~\cite{33},
and are briefly recalled in Sec.~\ref{App3} of the Supplemental Material~\cite{supX}.
A $\pi$-$\sigma$ mode is also present at energies larger than
 ${\sim}4$~eV~(Fig~\ref{F4comp2}).
\section{Loss-peak of the mixed phase and $\pi$-like plasmon
of {\Sio} silicene\label{secIII}}
Of particular significance, in the loss function analysis presented so far,
are the low energy features at $\omega{=}0.8{\myhyp}1.3$~eV,
for {\SiI} silicene, and $\omega{=}0.4{\myhyp}0.9$~eV,
for {\SiII}, {\SiIII} silicene, because they appear  not to be completely
erased by the interaction with Ag(111).

Indeed, the experimental loss of the pure {\AgII}
phase~[spectrum (ii) in Fig.~\ref{F2spec}] shows a peak,
which has some match with the low-energy end of
the theoretical loss function of {\SiII} silicene,
lying within the energy window of vertical transitions
associated with the $\pi$-$\sigma$ plasmon~(Figs.~\ref{F3bands}
and~\ref{F4comp}).

The interaction with silver destroys the high-energy loss properties
of the peeled phase and distorts the low-energy peak, which we
ascribe to a hybridized Si-Ag plasmon.
This interpretation is confirmed by existing DFT
calculations~\cite{31} and validated by our analysis on the {\AgII}
phase in Secs.~\ref{App1} and~\ref{App2} of the Supplemental Material~\cite{supX},
suggesting that the low-energy mode arises due to charge-density
oscillations within the silicene and first silver layer and keeps trace
of the peeled phase~(Figs.~\ref{s2} and~\ref{s3} in the Supplemental Material~\cite{supX}).
On the contrary, no signature of an interband plasmon
at $\omega{=}1.7{\myhyp}2{\,}{\rm{eV}}$ appears in the theoretical loss
spectrum of the silicene superstructures.

\begin{figure}[!!h]
\centering{\includegraphics[width=0.425\textwidth]{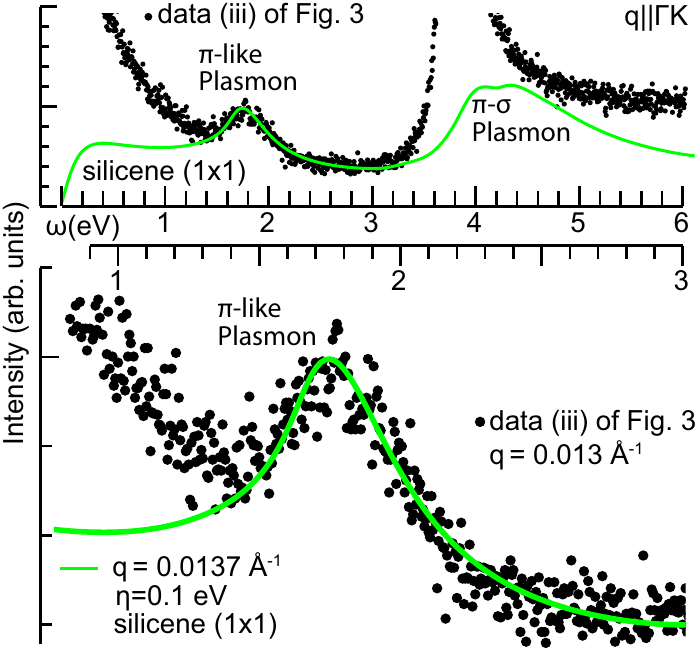}}
\vskip -6pt
\caption{Experimental loss spectrum (iii) of ${\rm{Fig.}}{\,}$~\ref{F2spec}
and theoretical loss function of {\Sio}
silicene, where the lowest sampled $q$ value along
${\Gamma{\rm{K}}}$ is considered with an overall broadening of
$0.1{\,}{\rm{eV}}$ to match the experimental resolution.}
\label{F4comp2}
\end{figure}

A more suggestive result is shown in Fig.~\ref{F4comp2},
where {\Sio} silicene calculations are compared with the
EL measurements on the coexisting silicene/Ag(111)
phases [loss-curve (iii) in Fig.~\ref{F2spec}].
The excellent agreement, at energy losses below $3{\,}{\rm{eV}}$,
lets us conclude that this morphology indeed presents a dielectric response
that matches what is expected for the {$\pi$-like}  plasmon of ideal silicene.
Also interesting to notice is that the $\pi{\myhyp}\sigma $ plasmon
 of {\Sio} silicene may be present as well, though it is hidden by
the Ag plasmon of the interface, as verified by comparing the loss
spectrum of the mixed phase
with that of clean Ag(111) [(i) in Fig.~\ref{F2spec}].

Hence, the interaction between the pure phases not only
weakens the hybridization of Si and Ag states,
but also appears to {\it merge} the different buckling levels
of supported silicene into a configuration that resembles
that of freestanding silicene.
Indeed, the different silicene superstructures are expected
to arrange into buckling conformations that mimic that of {\Sio}
silicene, when their distances relative to the Ag substrate
are increased above the $2.13{\myhyp}2.16$ {\AA} values,
predicted for the pure phases.
Our DFT calculations on the {\AgII} phase indicate that the {\Sio} silicene
domains begin to form for average Si-Ag distances larger
than $3.3$ {\AA} [see Fig.~\ref{scc} in Sec.~\ref{App1} of the
Supplemental Material~\cite{supX}].
This is the main result of the present study,
which awaits further scrutiny, especially regarding
the  ${\rm{low}}{\myhyp}q$ dispersion of the newly
discovered {$\pi$-like} mode [see Appendix~\ref{App3} of the
Supplemental Material~\cite{supX}].
\section{Conclusions\label{secIV}}
To summarize our results, we have presented a combined
experimental~(EELS) and theoretical~(TDDFT) approach
to show that silicene grown in a mixed phase on Ag(111)
preserves at least part of the semimetallic character of
its freestanding form, exhibiting an interband {$\pi$-like} plasmon
assisted by excitation processes at the M-point of the {\fBZ}.
Such a mode parallels the $\pi$ plasmon of freestanding graphene~\cite{33}.
The presence of a {$\pi$-like} plasmon by itself
does not allow us to conclude that the silicene overlayer maintains
its natural Dirac-cone structure.
For example, the interfaces of graphene on the (111) faces of copper
and nickel have a well-characterized $\pi{\myhyp}{\rm{plasmon}}$,
with the Dirac cone being preserved and destroyed, respectively,
by hybridization with the $d$ bands of the supporting metal~\cite{33aa,33bb}.
Nevertheless, in the latter case the shape, position and dispersion of the
plasmon peak is substantially different from that of freestanding
graphene~\cite{33cc}, which encourages performing further
investigations on mixed-phase morphologies of silicene on silver.
A similar approach would help to characterize the first VB and CB of
Silicene on Au(111)~\cite{47}, and corroborate the expected Dirac
cone properties of the system.

On the other hand, recent progress on the epitaxial synthesis of pure,
single phases of silicene on Ag(111), supported by DFT computations
 of their electronic properties, indicate that strong hybridization effects
make it impractical to separate a silicene structure with a well-defined
quasimetallic character.
An experimental fingerprint of this phenomenon is the hybridized Si-Ag plasmon,
discovered at low energy.
Our proposal is then to shift the efforts of making silicene a
feasible two-dimensional nanomaterial beyond graphene on the
epitaxial growth of silicene in mixed domains.
\section*{Acknoledgements}
A. S. acknowledges the computing facilities provided by the  \href{http://www.cineca.it/}{CINECA Consortium},  within the INF16\_npqcd project, under the
  \href{http://www.hpc.cineca.it/news/framework-collaboration-agreement-signed-between-cineca-and-infn}{CINECA-INFN}
 agreement, and the IsC49\_PPGRBGI project.

\setcounter{figure}{0}
\setcounter{table}{0}
\setcounter{equation}{0}
\setcounter{section}{0}
\renewcommand{\thesection}{S\arabic{section}}
\renewcommand{\theequation}{E\arabic{equation}}
\renewcommand{\thefigure}{F\arabic{figure}}
\renewcommand{\thetable}{T\arabic{table}}
\section*{Supplemental Material\label{SupM}}

\section{Electronic structure of the {\AgII} phase\label{App1}}
\begin{figure}[!!h]
\centering{\ \includegraphics[width=0.4\textwidth]{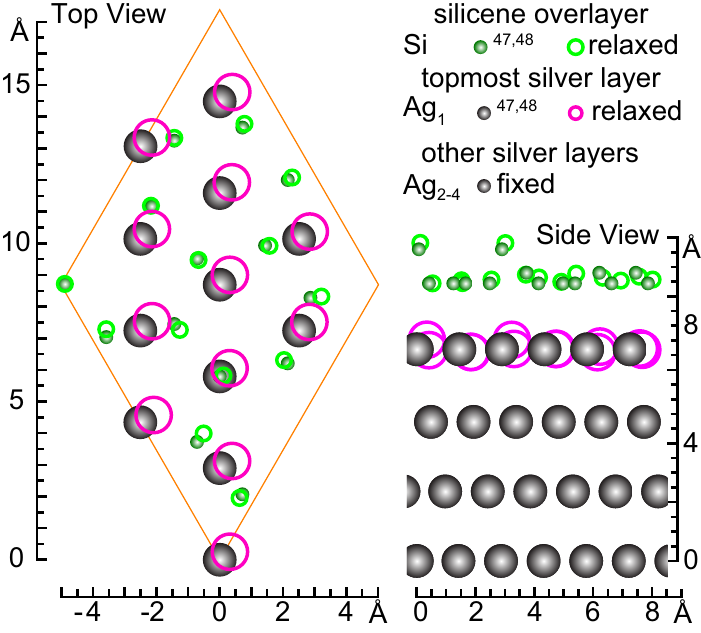}}
\vskip -6pt
\caption{Initial~\cite{27,28} and relaxed coordinates of the Si and Ag atoms in the {\AgII} phase.}
\label{s1}
\end{figure}

As mentioned in the main text, the positions of the Si atoms of the {\SiI}, {\SiII} 
and {\SiIII} silicene superstructures were computed by placing them on top 
of four silver layers of Ag in the {\AgI}, {\AgII} and {\AgIII} phases, respectively.
The optimized configurations of Fig.~\ref{F1phases} were used as inputs of the geometry optimization runs, performed using the Perdew-Burke-Ernzerhof~(PBE) exchange-correlation functional~\cite{36a}, in which coordinates of all Si and Ag atoms, of the topmost and second layers, respectively, were relaxed, using a $\Gamma$-centered MP grid~\cite{39} of $12{\times}12{\times}1$ wavevectors in the {\fBZ}.
In particular, for silicene {\SiII} on Ag {\AgII}, the lowest set of $302$ bands (with $292$ occupied)  were included in the calculation, yielding the relaxed coordinates shown in Fig.~{\ref{s1}}.

\begin{figure}[t]
\vskip -10pt
\centering{\includegraphics[width=0.38\textwidth]{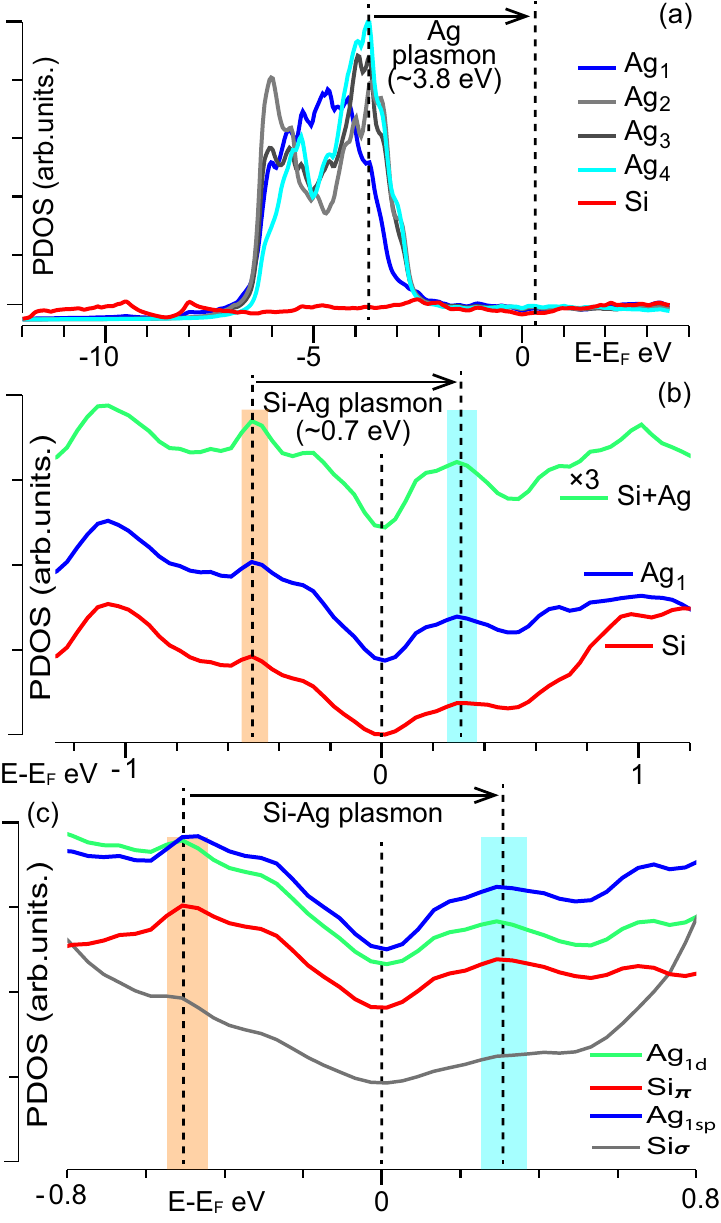} }
\vskip -6pt
\caption{Projected DOS of {\AgII} phase onto the Si and Ag atoms of the four layers considered in the present study. The shaded areas in Fig.~{\ref{s3a}} involve excitation processes within the peak width of the EELS spectrum in Fig.~{\ref{s3}}.}
\label{s2}
\end{figure}

Subsequently, Single-point self-consistent calculations were performed on the optimized Si/Ag structure with an MP grid of $24{\times}24{\times}1$ wavevectors, including the lowest set of $352$ bands.
The resulting DOS, projected~(P) onto the Si atoms of the {\SiII} superstructure and the Ag layers of the {\AgII} phase, is reported in Fig.~{\ref{s2}}.
As expected, the PDOS is dominated by intense peaks below ${\sim}3$~eV, generated by the weakly dispersing $d$ bands of silver; the black arrow line in Fig.~{\ref{s1}}(a) suggests that their contribution to the EL spectrum~(ii) of Fig.~\ref{F2spec} must be sought at energies around and above the Ag plasmon peak.

On the other hand, the lowest (occupied and empty) PDOS peaks, due to the topmost silicene and first Ag layers~[Fig.~{\ref{s1}}(b)], are consistent with the loss features at ${\sim}0.7$ eV of the data~(shown separately from other loss data in a zoomed scale in Fig.~{\ref{s3}}, below).
The atomic content of these peaks is mostly $\pi$ for the Si atoms and $\sigma$, $d$ for the Ag atom of the first slab layer~[Fig.~{\ref{s1}}(b)], in agreement with recent DFT calculations~\cite{31}.
Hence, the {$\pi$-like} symmetry of the bands in the peeled phase is lost due to hybridization, which is likely to produce a hybridized $s{\myhyp}p$ plasmon.

Other PDOS peaks, expected to play a role in the EL function at absorbed energies ranging from ${\sim}0.7$ to ${\sim}3.8$ eV, do not correspond to any significant variation in the same EELS spectrum, which follows a smooth trend from the hybridized Si-Ag to the Ag plasmon resonances.
We therefore expect that these PDOS structures, reported also in previous calculations~\cite{30}, are more and more smoothed by increasing the number of Ag layers, i.e.,  the number of $d$ bands,  in the slab system.

\begin{figure}[!!h]
\centering{\ \includegraphics[width=0.49\textwidth]{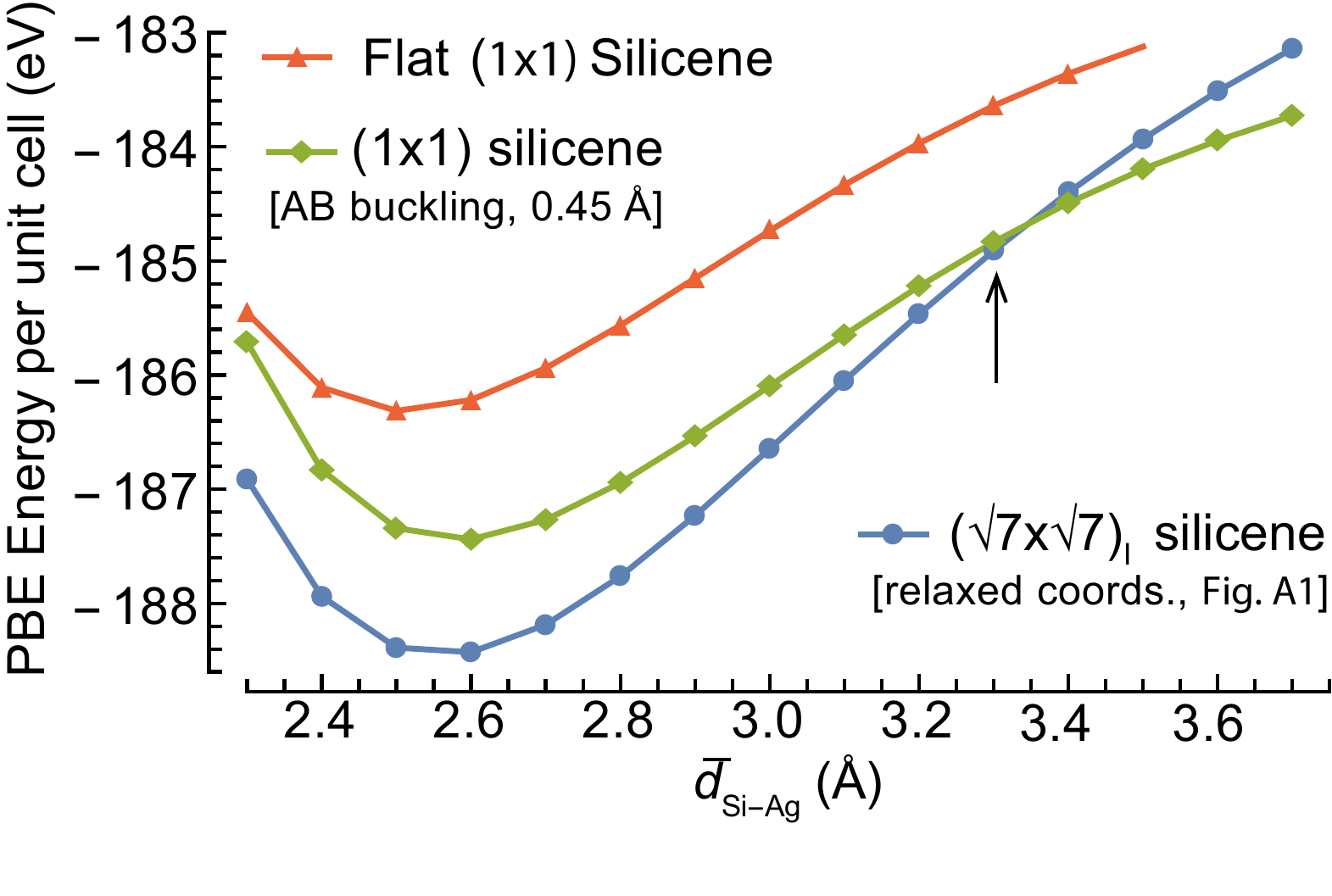}}
\vskip -6pt
\caption{Equilibrium DFT energy for {\SiII}, buckled {\Sio} and flat {\Sio} silicene at an average distance $\bar{d}_{\rm{Si-Ag}}$ from a four-layer slab of silver [see Fig.~\ref{s1}].}
\label{scc}
\end{figure}

A  more accurate insight into the role of hybridization comes from the 
analysis of the formation energy of the relaxed {\AgII} phase, as function 
of the distance between the {\SiII} silicene overlayer and the silver slab. 
In Fig.~\ref{scc}, we compare the total (PBE) energy per units cell required 
to have {\SiII} silicene and {\Sio} silicene (either in flat or buckled form) 
at an average distance $\bar{d}_{\rm{Si-Ag}}$  from the topmost silver layer. 
We see that {\SiII} silicene is energetically favored 
for $\bar{d}_{\rm{Si-Ag}}< 3.3$~{\AA}, whereas de-hybridized, 
buckled {\Sio} silicene is more likely to be formed 
for $\bar{d}_{\rm{Si-Ag}}> 3.3$~{\AA}.

Thus, the reduced hybridization observed in the mixed phase, 
and the appearance of a plasmon structure compatible with
 {\Sio} silicene, is consistent with an average separation larger 
than ${\sim}3$~{\AA} of the three silicene domains relative the
 topmost silver layer. 
Interestingly enough, a similar value characterizes the binding of 
graphene on Cu(111), being a major factor for the survival of the 
Dirac-cone structure of the interface.

\section{Loss function of {\SiII}  silicene and EELS spectrum 
of the {\AgII} phase \label{App2}}
To distinguish the delicate features of the $\pi$-$\sigma$ plasmon in {\SiII} 
silicene, we  discuss the energy-momentum features of the loss functions 
of the system in the energy range between $0.59$ and $0.96$ eV, 
where one-electron processes between occupied/empty bands of 
dominant $\pi$ and $\sigma$ character are exclusively 
involved~[see Fig.~\ref{Fxdisp}(b) and related discussion].

\begin{figure}[!!h]
\vskip -10pt
\centering{\includegraphics[width=0.495\textwidth]{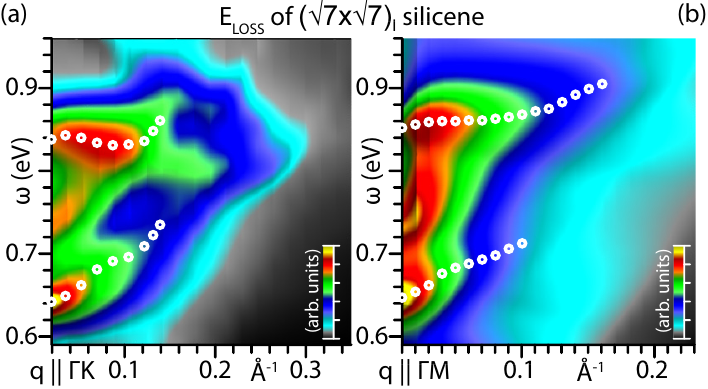} }
\vskip -6pt
\caption{Theoretical EL functions of {\SiII} silicene, computed with the
TDFT+RPA machinery illustrated in the main text, in response to an energy transfer $\omega$ in the range of $0.59$ to $0.96$~eV, and a momentum transfer below $0.33$~{\AA}$^{-1}$, along the $\Gamma{K}$ [(a)] and $\Gamma{M}$ [(b)] directions of the \fBZ [inset in Fig.~\ref{F3bands}(b)]. The loss-peaks are spotted as white circles.}
\label{s3a}
\end{figure}

In Fig~\ref{s3a} we display the anisotropic dispersions of this mode along 
the ${\Gamma}K$ and ${\Gamma}M$ paths of the reciprocal 
space~[Fig.~\ref{F1phases}(e) and inset in Fig.~\ref{F3bands}(b)].
When the Ag substrate, and its interaction with the {\SiII} silicene layer,
 is included in the calculations, the $\pi$-$\sigma$ mode is expected 
to survive, though being strongly contaminated by the aforementioned 
Si-Ag hybridization.

\begin{figure}[!!h]
\vskip -10pt
\centering{\ \includegraphics[width=0.38\textwidth]{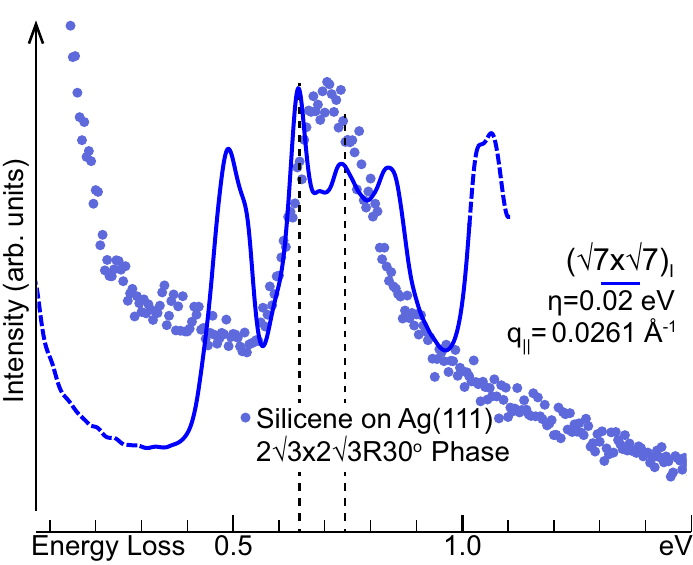} }
\vskip -6pt
\caption{Theoretical EL function of {\SiII} silicene and experimental loss spectrum (ii) of Fig.~\ref{F2spec}.
The lowest simulated value of $q$ is applied to the theoretical
loss curve, with a lifetime broadening of
0.02~eV [in Eq.~(\ref{AdlWi})].
}
\label{s3}
\end{figure}

In support of this idea, the EELS data, labelled (ii) in Fig.~\ref{F2spec}, 
are juxtaposed in Fig.~{\ref{s3}} to the theoretical EL curve of the {\SiII} 
superstructure. As noticed in the main text, the experimental low-energy 
peak at $0.75$~eV falls within the same energy-window of vertical 
transitions from the second VB to the first CB of {\SiII} 
silicene~ [Fig.~\ref{F3bands}(b)].
When Ag is included in the calculations, the electronic properties 
of the silicene layer change dramatically with respect to the peeled case, 
destroying the loss features at absorbed energies above ${\sim}1$~eV.

On the other hand, a comparison of the DOS of the {\SiII} superstructure 
alone~[Fig.~\ref{F3bands}(b)] with the different PDOS curves of the {\AgII} 
phase~(Fig.~{\ref{s3}}) indicates that the low energy features associated 
to the band levels, at energies below ${\sim}1$ eV from $E_F$, 
are distorted but not completely erased. 
Indeed, the shapes and relative weights of the PDOS of the {\SiII} overlayer 
and the first Ag layer at these energies are similar, whereas other Ag layers 
produce a constant background.

This let us evince that the $sp$ hybridization keeps traces of the peeled phase.
More accurate studies are, though, needed to put a final word on this interpretation.

\section{{$\pi$-like} plasmon mode of {\Sio} silicene vs 
EELS spectra of the mixed phase\label{App3}}
\begin{figure}[!h]
\centering{\includegraphics[width=0.495\textwidth]{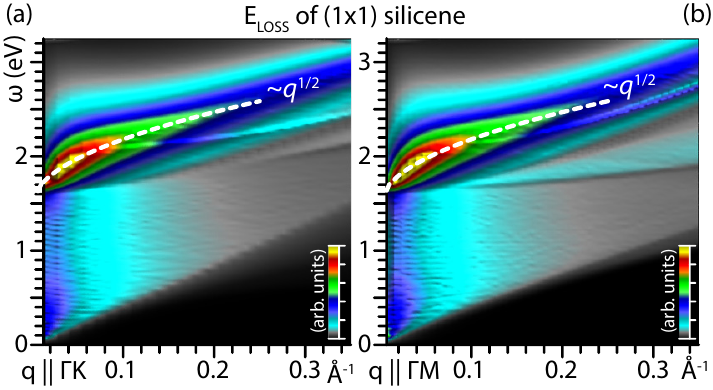} }
\vskip -6pt
\caption{Theoretical EL functions of {\Sio} silicene, computed with the
TDFT+RPA machinery illustrated in the main text, in response to an energy transfer $\omega$ below ${\sim}4$~eV, and a momentum transfer below $0.4$~{\AA}$^{-1}$, along the $\Gamma{K}$ [(a)]  and $\Gamma{M}$ [(b)] directions of the {\fBZ} of Fig.~\ref{F3bands}(d). The square-root-like low-momentum dispersions are shown by white dashed lines.}
\label{s4A}
\end{figure}

The 2D plasmon nature of the {$\pi$-like} plasmon of {\Sio} silicene has 
been scrutinized in a previous study~\cite{33}, 
where the effect of the $sp^2$-$sp^3$ hybridization has 
been carefully addressed. 
Indeed, as shown in Fig.~\ref{s4A}, the low-energy and low-momentum 
properties of the loss functions of the system describe a perfect two-dimensional 
mode with square-root-like dispersion along 
both ${\Gamma{K}}$ and ${\Gamma{M}}$.

\begin{figure}[t]
\centering{\includegraphics[width=0.44\textwidth]{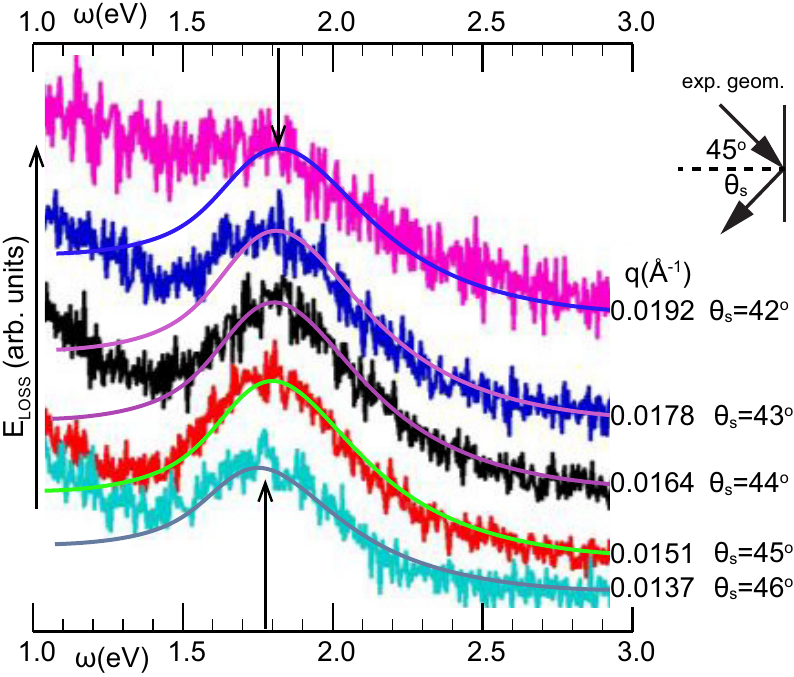} }
\caption{EELS spectrum acquired from silicene on the mixed phase with the LEED structure of Fig.~\ref{leedx}(b) and theoretical EL function at small in-plane momenta in the range of $0.013$ to $0.020$ {\AA}$^{-1}$.}
\label{s4}
\end{figure}

On the experimental side, Fig~{\ref{s4}} shows EELS measurements on the mixed 
phase performed with the same geometry used for all the spectra of 
Fig.~{\ref{F2spec}} at different scattering angles, obtained by rotating 
the analyzer in the range of $42^{\circ}$ to $46^{\circ}$ 
from the perpendicular direction to the sample.

What is apparently strange here, is that the frequency of the {$\pi$-like} 
mode is subject to small variations, of the order of $1\%$, vs significant 
changes in the scattering angle. 
These would correspond to in-plane momentum variations in the
 range of $0.012$ to $0.18$~{\AA}$^{-1}$, as predicted by the conservation 
of kinetic energy and in-plane momentum
$$q{=}\sqrt{2E_{i}}\sin \theta _{i}-\sqrt{2(E_{i}-\omega_p)}
\sin \theta _{s},$$ 
typically used in EELS experiments, with $\omega_p$=1.75-1.83 eV being 
the lower loss peak resonance.

The quasi dispersionless behavior observed here is carefully related to the 
morphological status of the silicene overlayer, i.e., to the mixed domains with 
juxtaposed diffraction patterns.

As recently demonstrated~\cite{poli}, truly dispersing plasmonic modes 
are achieved only in surfaces exhibiting large-scale, flat terraces. 
On the other hand, surfaces with extensive large-scale faceting and shattering 
are characterized by a nearly constant behavior of the plasmon dispersion, 
across the whole {\fBZ}.
In our case, the epitaxial growth of multiphase silicene inherently poses 
limits on the crystalline quality of the overlayer, which, owing to the different 
morphologies of the coexisting silicene superstructures, cannot possibly exhibit 
large-scale, flat terraces as single crystals of layered materials exfoliated with 
precision methods. 
This is confirmed upon comparing the LEED spots of Fig.~\ref{leedx}(b) 
and the mixed reciprocal space of the {\AgI}+{\AgII}+{\AgIII}
 phase of Fig.~\ref{leedx}(c), as suggested in the main text.

In typical EELS setups, the impinging electrons from the primary beam are reflected inside an extended electron-density distribution at the surface~\cite{naza}.
In the presence of surface corrugation, these electrons are also reflected by the crests of the rough surface.
Hence, the specular reflection of impinging electrons is inevitably lost due surface roughness.
The experimental fingerprint of this phenomenon is the missing dependence of the intensity of reflected electrons on the scattering angle, which we have experienced in our EELS experiments.
After scattering from the sample, the electrons are diffused by the crests of the rough surface in a wide solid angle. The corresponding excitation spectrum experimentally probed by EELS has then momentum-integrated spectral contributions with a maximum intensity due to low momenta. Thus the spectra of Fig.~{\ref{s4}} are not resolved in in-plane momentum, but they are rather dominated by low-momentum excitations.

We may tentatively attribute a leading parallel momentum to the spectra, based on the theoretical loss curves computed from {\Sio} silicene with the TDDFT+RPA approach of the present work.
As also shown in Fig.~{\ref{s4}} the agreement is significant in all considered cases, which let us think that indeed the plasmon structure that we are observing is due to a quasi freestanding silicene overlayer. 
\end{document}